\documentclass[fleqn,10pt]{wlscirep}
\usepackage[utf8]{inputenc}
\usepackage[T1]{fontenc}

\usepackage{lineno}

\usepackage{booktabs}
\usepackage{multirow}
\usepackage{longtable}
\usepackage{multicol}
\usepackage{threeparttable}
\usepackage{tabularx}
\usepackage{array}

\newcolumntype{Y}{>{\raggedright\arraybackslash}X}

\makeatletter
\newcommand{\beginsupplementaryinformation}{%
  \clearpage
  \newgeometry{left=3cm,right=3cm,top=3cm,bottom=3cm}%
  \pagestyle{plain}%
  \setlength{\parindent}{15pt}%
  \setlength{\parskip}{0pt}%
  \renewcommand{\rmdefault}{ptm}%
  \normalfont
  \fontsize{11}{13.6}\selectfont
  \captionsetup{labelfont=bf,textfont=normalfont,labelsep=period,justification=justified,singlelinecheck=false}%
  \setcounter{figure}{0}%
  \setcounter{table}{0}%
  \renewcommand{\thefigure}{S\arabic{figure}}%
  \renewcommand{\thetable}{S\arabic{table}}%
  \setcounter{section}{0}%
  \setcounter{subsection}{0}%
  \setcounter{subsubsection}{0}%
  \titleformat{\section}{\normalfont\Large\bfseries}{\thesection}{0.75em}{##1}%
  \titleformat{name=\section,numberless}{\normalfont\Large\bfseries}{}{0em}{##1}%
  \titleformat{\subsection}{\normalfont\large\bfseries}{\thesubsection}{0.75em}{##1}%
  \titleformat{\subsubsection}{\normalfont\normalsize\itshape}{\thesubsubsection}{0.75em}{##1}%
  \titleformat{\paragraph}[runin]{\normalfont\normalsize\bfseries}{}{0pt}{##1}%
  \titlespacing*{\section}{0pt}{3.5ex plus 1ex minus .2ex}{2.3ex plus .2ex}%
  \titlespacing*{\subsection}{0pt}{3.25ex plus 1ex minus .2ex}{1.5ex plus .2ex}%
  \titlespacing*{\subsubsection}{0pt}{3.25ex plus 1ex minus .2ex}{1.5ex plus .2ex}%
  \titlespacing*{\paragraph}{0pt}{1ex}{1em}%
}

\newcommand{\supplementarytableofcontents}{%
  {\centering\large\bfseries Supplementary Notes\par}%
  \vspace{0.75\baselineskip}%
  \@starttoc{stc}%
}
\makeatother

\title{%
No one likes it hot, but hotter cities adjust by staying active later
}
\author[ab,*]{Andrew Renninger}
\author[c]{Till Koebe}
\author[c]{Ingmar Weber}

\affil[a]{Department of Network and Data Science, Central European University}
\affil[b]{School of Geographical \& Earth Sciences, University of Glasgow}
\affil[c]{Saarland Informatics Campus, Saarland University}
\affil[*]{Corresponding author: Andrew Renninger (E-mail: renningera@ceu.edu)}

\begin{abstract}
Extreme heat suppresses urban activity, but its effects need not be uniform across climates or across the day. Using data on activity at points of interest in 20 cities spanning temperate, tropical, and arid environments, we show that hot days reduce activity overall while shifting it away from midday and toward later hours. This rescheduling is substantially stronger in historically hotter cities, which exhibit smaller losses and larger evening substitution. To understand these changes, we introduce a Bactrian index of bimodality, which measures the degree to which a city’s daily activity profile has one hump or two---one during the day and another during the evening. Arid desert cities like Doha, Amman, and Kuwait City are more Bactrian in level, but cities like Milan become Bactrian on hot days. Together, our results suggest that adaptation to heat in cities operates less through avoiding activity altogether than through moving it to cooler hours. This provides channels for adaptation in cooler cities, but it also suggests limits to adaptation in warmer ones: as evenings become warmer, these too may become intolerable. 
\end{abstract}

\begin{document}

\flushbottom
\maketitle

% \begin{linenumbers}

\section*{Introduction}
% INTRODUCTION
Extreme heat is becoming more frequent, intense, and socially consequential in a warming climate \cite{zhang2015impact, schaffer2012emergency, perkins2012increasing, perkins2020increasing}. Because ambient temperature interacts with the built and natural environments, urban areas are more exposed to extreme heat \cite{}. Further, cities today both require outdoor activity---such as construction---and many, in developing countries in particular, lack adequate indoor cooling \cite{}. This means that adaptation to heat depends not just on infrastructure and engineering but on social and behavioural solutions as well \cite{turek2021adaptation, navasmartin2024framework}. In the following study we explore common adaptation strategies across diverse climates by examining activity patterns in 20 cities from arid, tropical and temperate zones. We find that delaying or rescheduling activity into the cooler hours is common, but that this strategy is more pronounced in arid desert cities. 

% CONTEXT
Extreme heat is increasingly a first-order constraint on economic performance, health, and everyday urban life. Macro evidence shows that temperature shocks depress growth in poorer economies and that aggregate production is nonlinear in climate even when rich and poor countries are pooled: output peaks around a moderate annual average temperature and declines sharply at higher temperatures \cite{dell2012temperature, burke2015production}. Heat threatens labour productivity: in Indian manufacturing, heat decreases worker productivity by as much as 9\% per degree and heat increases worker absences \cite{somanathan2021productivity}, and cooler lighting on a factory floor raises productivity \cite{}; in Chinese manufacturing, plant data reveal a curved relationship between temperature and productivity, with benefits at warm temperatures and costs at hot ones, implying large losses under future warming \cite{zhang2018reallocation}. Heat also influences cognition: hotter school years reduce learning without air conditioning \cite{park2020learning}, and \cite{adhvaryu2020light}. This pattern is consequential because it is not confined to outdoor or obviously heat-exposed sectors. Heat reaches physical and cognitive tasks, indoor and outdoor settings, and both labor supply and on-the-job effectiveness. Heat is not only a health shock, raising mortality and morbidity \cite{vicedocabrera2021mortality}, but a constraint on labor and cognition that can propagate from workers and firms to aggregate production.

Research also points to local adaptation. While extreme heat is responsible for a rising number of deaths each year \cite{zhang2015impact}, in both milder climates where heat waves are rarer \cite{zhao2021mortality, vicedocabrera2021mortality} and hotter areas where they are common \cite{mazdiyasni2017increasing}; the most frequent temperature in a region is typically associated with the fewest deaths \cite{yin2019mmt}, which suggests that populations learn to live with common temperatures. 

Behavioural evidence shows that one important adjustment margin is neither full avoidance nor full exposure, but reallocation of activity across time and space. High temperatures reduce hours worked in climate-exposed industries and outdoor leisure in U.S. time-use data, while shifting time toward indoor leisure, consistent with short-run temporal substitution \cite{graffzivin2014time}. New mobility datasets reveal the same logic more directly. In Houston, extreme heat suppresses daytime movement and raises the share of activity after 8 p.m., with delayed evening travel times and substantial spatial heterogeneity across neighborhoods \cite{tian2024mobility}. In the US, heat waves shift time use, with some groups showing much weaker capacity to adapt than others \cite{batur2024mobility}. Across five Chinese cities, physical activity peaks at about 16--19.3\textdegree C in temperate settings and falls by roughly 800--1500 daily steps at higher temperatures, whereas declines are muted and statistically weak in subtropical Shenzhen and Hong Kong \cite{ho2022activity}. Heat thus changes not only how much activity occurs, but when it occurs and whether it can be shifted toward cooler or more protected environments. 

Research on urban heat adaptation often emphasises the spatial margin: shade, vegetation and cooling \cite{turek2021adaptation,bell2024heat}. A parallel literature, however, points to the temporal margin. Exposed residents and workers respond to extreme heat by shifting work or errands to cooler hours, or substituting away from the hottest part of the day \cite{graffzivin2014time, lopalo2023worker, parsons2021labor}. Conceptually, we can think of the spatial margin as capital-heavy and the temporal margin as capital-light, which makes it attractive in settings with limited state resources for subsidies or investment---but it requires coordination and cooperation. Opening hours may need to adjust, for example, and transit agencies may need to redistribute services over the day. 

% GAP
A growing body of evidence supports this idea. Outdoor leisure activity in China shifts away from noon and late afternoon toward cooler morning and evening hours on hot days \cite{fan2023intraday}. Mobile-phone data from the San Francisco Bay Area and Houston show that higher temperatures reduce mobility and can delay activity toward later hours \cite{ly2023mobility, tian2024mobility}. Recent transaction data from Australian cities similarly show that spending collapses during the hottest afternoon hours on $\geq 35^\circ$ C days while later-day activity is more resilient \cite{seijas2026adaptive}. These studies are important, but they are mostly single-country or single-region analyses, and they often do not distinguish clearly between two questions: whether extreme heat reduces total urban activity, and whether it changes \emph{when} activity occurs.

An indication that hot cities respond to heat in different ways emerges from triangulating evidence from across countries. For example, in Spain heat both reduces and retimes activity \cite{renninger2025extreme}, but in Singapore, hotter days increase rather than decrease visits to offices, malls, parks, and schools with air conditioning: a 1\textdegree C rise in the daily maximum heat index raises office visits by $\sim 0.4\%$, mall visits by up to $\sim 0.6\%$, park visits by $\sim 0.9\%$, and commuting to schools by 2.6\%, with especially strong responses later in the day when temperatures crest \cite{fesselmeyer2024activity}. This suggests not only that cooling infrastructure plays a critical role in adaptation, but that its utilisation during heat waves requires behavioural adjustments: if cities in extreme climates were behaviourally unaffected by heat, we should observe a null effect of heat on mobility on hot days. 

% MOTIVATION
Literature points to clear adaptations---reducing, retiming, and relocating activity. Yet while the public health literature points to evidence for adaptations, the microfoundations of these adaptations are less clear. Although many studies document the shape of urban activity across the day \cite{renninger2025extreme, li2022spatiotemporal, albouy2016climate, graffzivin2014time} and ask whether hot places differ from cooler places in level, timing, and substitution patterns. Existing evidence across countries is mixed: India shows clear retiming in mobility data but Mexico does not \cite{renninger2024remote}. In the following paper we address this gap by comparing across cities with different climatic contexts. We merge data on weather with measures of urban ``busyness'' at points of interest in 20 cities from April 2021 through April 2023 and explore differences in responses to hot days across historically cooler and hotter cities. Our primary contribution is substantive: the data indicate that extreme heat changes \emph{when} urban life happens more than \emph{whether} it happens, although activity does fall on aggregate across our sample. A secondary contribution is comparative: hot cities shift activity later in the day relative to other cities, and we document a ``Bactrian'' shape---two humps of activity, at midday and in the evening, rather than one peak---that is more pronounced in hot cities than mild ones, indicating adaptive redistribution. Our results suggest both adaptive opportunities and limits, as rising temperatures make evenings too hot for intertemporal substitution. 

\begin{figure*}[ht!]
\centering
\includegraphics[width=1\textwidth]{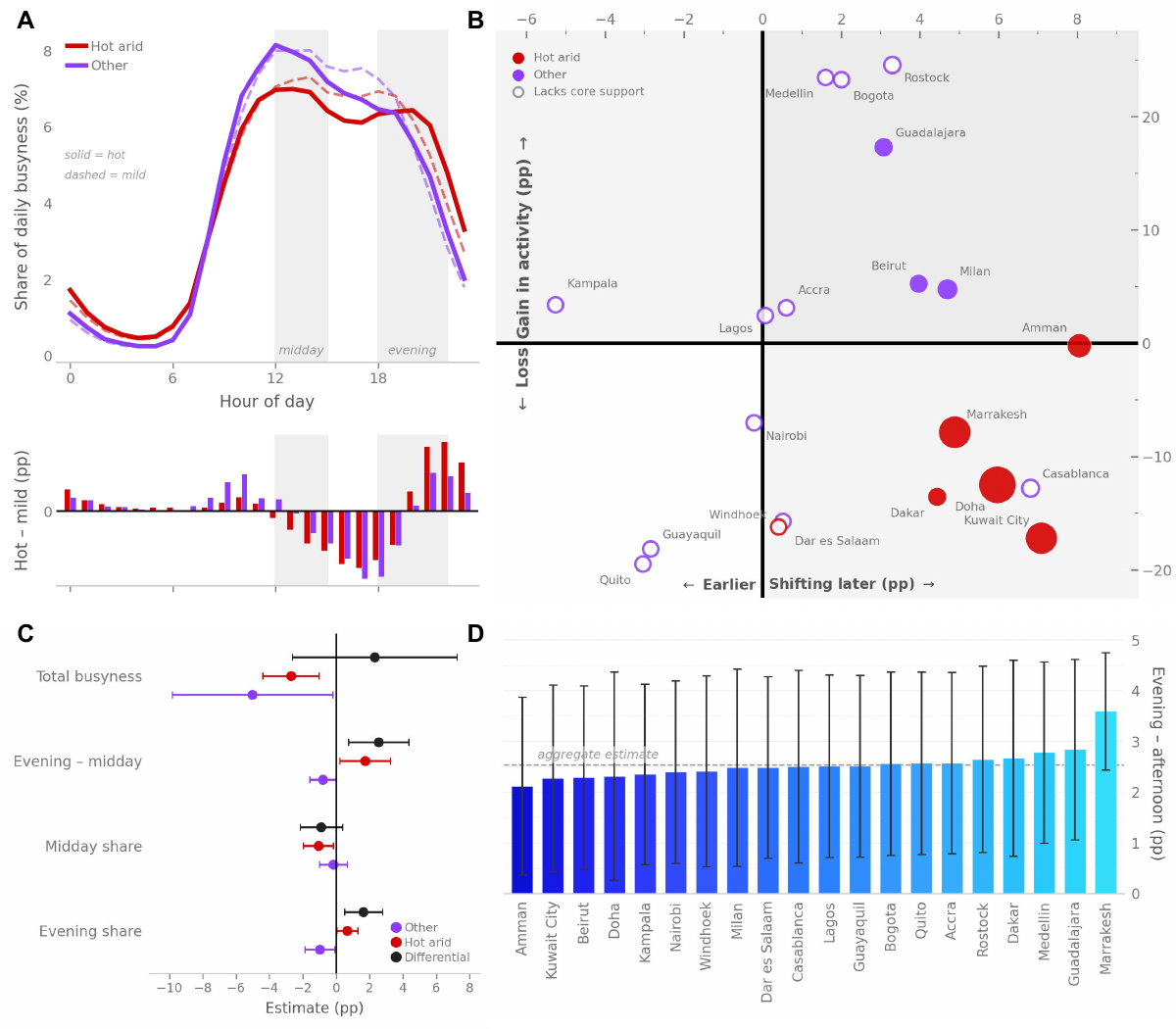}
\caption{\textbf{Hotter cities stay active later.} \textbf{A} Grouped activity profiles on locally hot and mild days, showing that cooler climates have a rise through the day and fall in the evening, while hotter ones have a second rise in the evening when temperatures fall; though subtle, we also see that hotter days show depressed activity in the afternoon and expanded activity in the evenings---especially in arid cities. \textbf{B} Loss or gain in activity on days where temperatures exceed 35\textdegree C, relative to days with temperatures below 25\textdegree C, against shifted activity on those days; we see that hot, arid cities see reduced but also retimed activity on hot days. \textbf{C} Estimated effects of temperatures exceeding 35\textdegree C on total busyness, evening-against-midday share, midday share, and evening share; modelled estimates again show both losses midday but gains in the evening for hot, arid cities, creating greater retiming on aggregate. \textbf{D} We use leave-one-city-out estimates to demonstrate the durability of our results, which are robust but also fall when we hold out hot, arid cities, indicating that these cities have the strongest responses to extreme heat.}
\label{fig1}
\end{figure*}

\section*{Results}
In order to understand how cities adapt to extreme heat, we distinguish between two margins of response: whether activity falls on the day---reducing---and whether it is rescheduled within the day, or retiming. We address these questions with complementary descriptive and modelled evidence, comparing hot and mild days across 20 cities observed from April 2020 through April 2023 and centering the analysis on $35$--$40^\circ$C days relative to $20$--$25^\circ$C days, where most cities have observations: hot, arid cities contribute 623 city-days in the $35$--$40^\circ$C bin, compared with 34 in all other cities, while above $40^\circ$C the sample is limited to only the desert cities. We therefore treat $35$--$40^\circ$C as the core comparison. (Note that in the following analysis, when a city lacks core support, look to changes between median and temperatures from the top third of the distribution for that city.) The twin goals of our study are to understand if heat suppresses activity, and if it shifts activity away from midday and toward later hours, with that adaptive change stronger in historically hotter cities.

We begin by documenting important regularities in the data. The first concerns the shape of the distribution of activity across the day: as we see in Fig. ~\ref{fig1}\textbf{A}, all cities show a rise in activity through the day, cresting in the afternoon, but hot and arid cities show another rise in activity in the evening as temperatures fall. Extreme heat also changes when activity occurs, relative to cooler days: in hot, arid cities, hotter days suppress activity the middle of the day and stimulate it later into the evening. Figure ~\ref{fig1}\textbf{B} shows that pattern for each city: cities in hotter climates tend to shift activity into the evening while reducing activity generally. Amman comes closest to a ``late, not less'' pattern, whereas Doha and Kuwait City shift strongly later but still lose activity overall. Milan is the clearest non-hot-arid case in which the clock bends toward the evening.

The modelled estimates in Fig.~\ref{fig1}\textbf{C} confirm that the stronger response is in timing rather than level changes. Relative to $20$--$25^\circ$C days, $35$--$40^\circ$C days are associated with a 5.0 percentage point decline in total busyness in non-hot-arid cities and a 2.7 percentage point decline in hot-arid cities. The clearer adjustment, however, is retiming. In non-hot-arid cities, $35$--$40^\circ$C days see a fall in the evening-minus-midday share by 0.8 percentage points; these same temperatures in hot-arid cities correspond to a rise of 1.7 percentage points, for a difference of 2.5 percentage points. This shift combines a larger midday decrease in hot-arid cities ($-1.07$ versus $-0.18$ percentage points) with an evening increase in hot-arid cities ($+0.66$ versus $-0.98$ percentage points).

The timing result is not driven by a single city. In leave-one-city-out checks, shown in Fig.~\ref{fig1}\textbf{D}, the differential in evening-minus-midday share remains positive in every run, ranging from 2.12 to 3.60 percentage points. The January--September 2022 window yields nearly identical timing estimates---2.28 percentage points with date fixed effects and 2.47 percentage points with seasonal controls---and implies a peak hour about 1.2 to 2.0 hours later in hot-arid cities on very hot days. Across specifications, then, hotter cities do not simply do less; they do later.

In Supplementary Fig. ~\ref{fig:si_sd_variation} we show that these results hold when using relative temperature anomalies for each city instead of the absolute temperature thresholds, which allows us to include more cities that lack the highest temperatures. Yet there is an informative exception to our results: tropical cities with little variation in temperature do not appear to reduce or retime activities---which is either the result of a lack of statistical power with limited variation, or acclimation. Using ``apparent'' temperature in Supplementary Fig. ~\ref{fig:si_apparent}, which considers wind, humidity and solar radiation to model how it \emph{feels}, we do see that humid tropical cities retime activity into the evening as much as, or more than, hot and arid ones; in this specification, we lose statistical power when we add a third category---other, tropical, arid---but we see similar point estimates and signs. The broader implication is that once heat is measured as people feel it, rescheduling into cooler hours emerges as a general response rather than an arid one.    

\begin{figure*}[t]
\centering
\includegraphics[width=1\textwidth]{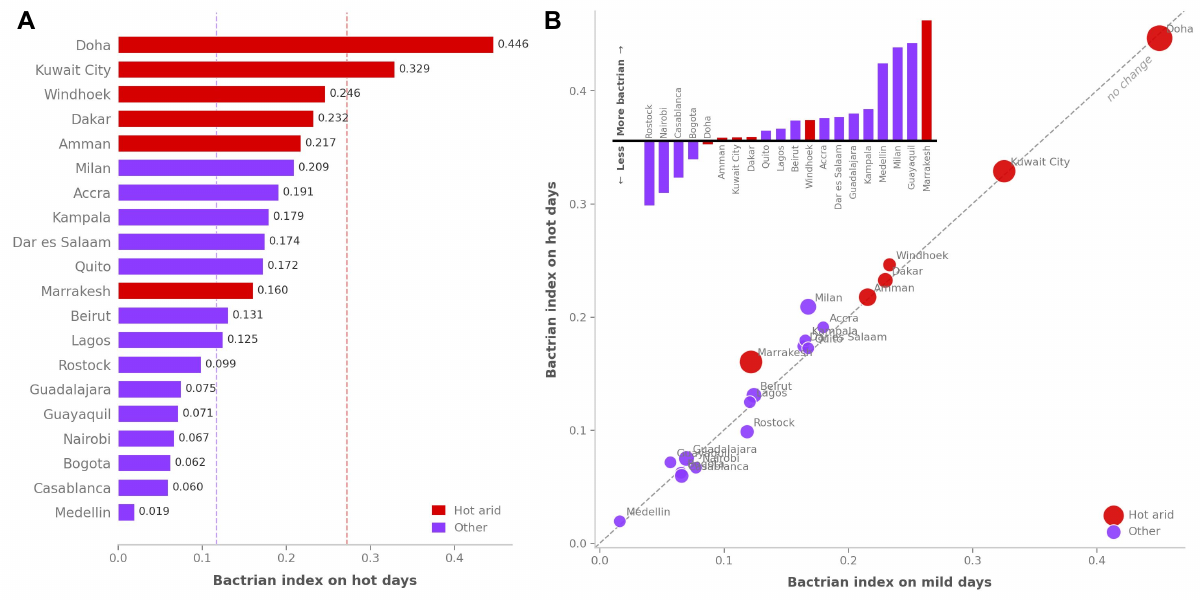}
\caption{\textbf{Hotter cities have multiple active periods.} \textbf{A} Our ``Bactrian'' index on locally hot days, ordered from highest to lowest, showing that hot, arid cities are also the most Bactrian---with multiple active periods; this index is defined as the amplitude of the second daily harmonic relative to the first. \textbf{B} Bactrian index on mild days and hot days for each city: hot and arid cities are more Bactrian on all days but other cities like Milan see the greatest shift toward Bactrian distributions between mild and hot days.}
\label{fig2}
\end{figure*}

Heat also changes the \emph{shape} of the urban day. Figure ~\ref{fig2}\textbf{A} ranks cities by a Bactrian index, defined as the amplitude of the second daily harmonic relative to the first; higher values indicate a more pronounced bimodal distribution, with one crest during the day and another in the evening. Hot and arid cities stand out on this measure, including Doha (0.446) and Kuwait City (0.329). Figure ~\ref{fig2}\textbf{B} shows that the key distinction is one of level rather than slope: that hot, arid cities typically have higher Bactrian values on both mild days and hot ones, indicating that fragmented schedule is part of normal urban rhythm in these places rather than a response that appears only under stress. The strongest changes instead occur in Milan ($+24.5\%$), Guayaquil ($+25.7\%$) and Marrakesh ($+31.6\%$), suggesting that cities outside the hottest climates see disruptions that cities inside them do not. In the supported core of the sample, the Bactrian index rises from 0.164 to 0.292 in hot, arid cities and from 0.093 to 0.188 in other cities.

Marrakesh stands out here as a city that is hot and arid while exhibiting a pronounced Bactrian shift on hot days. We treat this as an illustrative exception to the rule: while Amman, Doha and Kuwait City have most frequent temperatures of $\sim30^\circ$C, $\sim39^\circ$C and $\sim45^\circ$C, respectively, Marrakesh's most frequent temperature is $\sim22^\circ$C, so while it is in the same desert category, it may not be as well adapted to heat as the cities that experience temperatures greater than $30^\circ$C on a regular basis. We also find an interesting case in Doha, which becomes \emph{less} Bactrian on hot days---but this occurs, as we see in Supplementary Fig. ~\ref{fig:city_profiles}, because the distribution collapses towards the evening. 

The move into later hours is not uniform across urban life. We disaggregate the analysis by place category to locate where activity changes under heat, and although these estimates remain exploratory because we lack statistical power at this level of detail, they point to a consistent pattern. The differential between hot, arid cities and others is largest in health venues (20.7 percentage points), groceries (9.4), and petrol stations (7.5), and essentially absent in shopping malls ($-1.4$). These estimates are noisier and should be read with caution, but they suggest that cities can postpone necessary or routine activities more easily than purely discretionary ones. 

On level, hot and arid cities generally exhibit smaller losses on hot days in total busyness across venue categories, but estimates per category are imprecise and do not isolate a single driver. The timing margin is clearer. Pooling across place types, hot and arid cities show a 4.9 percentage point larger increase in evening-minus-midday activity share on very hot days, and some category estimates are positive: the clearest interpretable cases appear in shopping malls and gas stations, with recreation also positive but imprecise, while food and dining is the main exception. Collapsing venues into broader bins sharpens that picture further, with larger retiming differentials for discretionary than routine destinations (about 8.0 versus 1.8 percentage points) and for more exposed than cooled indoor settings (about 7.4 versus 2.4 percentage points). As shown in Supplementary Fig. ~\ref{fig:si_shiftshare}, these patterns do not appear to be driven by differences in amenity composition: a shift-share decomposition assigns the aggregate retiming gap primarily to changes to visitation patterns within categories, while differences attributable to composition are small and, if anything, offsetting.

Taken together, the results show that heat does suppress urban activity, but the more distinctive adjustment---especially in places long exposed to severe heat---is temporal. Cities do not simply empty out. They become quieter at midday, busier later, and more double-peaked across the day.

\section*{Discussion}
Our results speak most to the value of learning from hot, arid cities. These places sit closest to the thermal frontier that many other cities may approach, and they show that adaptation to heat is not the same thing as immunity from heat. Very hot days still reduce activity there. What differs is the margin of adjustment: relative to other cities, hot-arid cities show smaller aggregate losses, stronger substitution away from midday and into the evening, and daily profiles that are more persistently double-peaked. In that sense, the most exposed cities in our sample are cases in which the social organisation of the day has already adapted to make some activity survivable under extreme conditions.

This matters because it reframes what urban heat adaptation looks like. Much of the literature and much policy discussion focus on the spatial margin of adaptation---shade, vegetation, surface and building materials, and cooling technologies. Our results point to a parallel temporal margin. The Bactrian pattern we document is not simply a curiosity of daily shape; it is evidence that cities reorganise activity across hours when midday becomes costly. The supplementary place category analysis sharpens that interpretation. Although the venue estimates are exploratory, the aggregate retiming differential does not appear to arise mainly because hot and arid cities contain different kinds of amenities. Rather, similar kinds of places are used at different times of day under heat. That suggests that adaptation is partly about infrastructure, but also about schedules, opening hours, transport provision, and the capacity of workers and consumers to coordinate around cooler windows.

Another compelling finding---or null result---in this study is the tropical anomaly. When we replace absolute thresholds with locally defined heat anomalies, the cities in our sample with limited variation---many of them tropical on the hot end or mountainous on the cool end---show no clear evidence of either activity loss or retiming. In tropical cities in particular, day-to-day temperature variation is often small, so hot days may simply not provide enough statistical leverage to evidence the kinds of responses that appear elsewhere. Yet the opposite interpretation is also possible: where heat exposure is chronic and expected, adaptation may already be so embedded in buildings, routines, or expectations that marginal shocks do not induce adaptive change. At present we cannot distinguish between these explanations, and that uncertainty is itself important. That we see this null effect on both hot and cool cities without strong variation suggests that this is not inoculation. 

Generally, our findings suggest that heat changes the timing and texture of urban life as much as its aggregate intensity. Cities do not simply empty out under heat. They become quieter at midday, busier later, and in many cases more segmented across the day. That distinction matters for how impacts are experienced. A city that preserves activity by shifting it later is more resilient than one that simply shuts down, but it is also reorganising work, consumption, care, and travel around a narrower thermal window. One implication is that adaptation policy should pay more attention to the governance of urban time, not only to the engineering of urban space.

Several limitations bound these conclusions. We observe relative busyness at points of interest rather than welfare, productivity, or individual exposure, so we can identify behavioural redistribution more readily than its costs and benefits. Our sharpest cross-climate comparison is necessarily concentrated in the 35--40$^\circ$C range, because above 40$^\circ$C only the desert cities provide meaningful support. And the place-type results remain suggestive rather than definitive. Still, the broad pattern is clear enough to matter: hotter cities do not simply do less; they do later. That adaptive margin may offer a path of adjustment for cities that are warming into more dangerous temperature ranges. It may also prove fragile. If evenings and nights become less tolerable, the temporal refuge that is visible in our data may begin to close, reducing one of the most accessible forms of behavioural adaptation now available.

\section*{Methods}
We measure urban activity using Google's Popular Times data, scraped in real time from the Google Places API across 20 cities on four continents over three years, from April 2020 to April 2023. For each of approximately 3.6 million observations---stratified by city, day, hour, and place category---Google reports a ``busyness'' score derived from aggregated, anonymised location signals from mobile users. The scores are aggregated across all venues of a given type within each city, so the absolute scale carries no meaning \emph{between cities}; we therefore construct all outcomes as ratios \emph{within cities}---either the share of a day's total activity falling in a given time window (for retiming) or the percentage deviation from an index value (for reducing). Google periodically recalibrates the index, which limits pandemic confounding, and these rebases are timed across cities; we also implement modelling strategies---like fixed effects---to absorb any other confounding. For each observation, we join temperature and precipitation, drawn from Open Meteo reanalysis data and matched at the level of the city and day.

With weather appended to busyness, we then model the effect of temperature on both the level and timing of activity. We restrict the sample to days falling in two absolute temperature bins---20--25\textdegree C (mild) and 35--40\textdegree C (hot)---which provides a clean contrast with common support across both hot-arid and non-hot-arid cities. Days outside these bins are excluded from estimation. We classify six cities as hot-arid (Doha, Amman, Dakar, Marrakesh, Windhoek, Kuwait City) and pool the remaining fourteen as ``other.'' (For a complete list of cities and their range of temperatures, see Supplementary Table ~\ref{T1}.) For both level and timing, we estimate a two-way fixed effects (TWFE) specification of the form
\[
y_{cdt} \;=\; \beta_1 \,\mathit{Hot}_{ct} \;+\; \beta_2 \,\mathit{Hot}_{ct}
\times \mathit{HotArid}_c \;+\; \gamma \,\mathit{Precip}_{ct} \;+\; \alpha_c
\;+\; \mu_m \;+\; \delta_d \;+\; \varepsilon_{cdt}
\]
where $\mathit{Hot}_{ct}$ is an indicator for days in the 35--40\textdegree C bin, $\alpha_c$ are city fixed effects, $\mu_m$ are year--month fixed effects, and $\delta_d$ are day-of-week fixed effects. Standard errors are clustered at the city level. Under this specification, $\beta_1$ captures the effect of heat on non-hot-arid cities, $\beta_1 + \beta_2$ gives the effect on hot-arid cities (recovered as a linear combination with covariance-adjusted standard errors), and $\beta_2$---the interaction---is the differential: how much more (or less) hot-arid cities respond to the same absolute heat.

We estimate this specification for four dependent variables: total busyness on the day (the percentage deviation from each city's index value), the midday share of daily activity (hours 12--15), the evening share (hours 18--22), and the evening-minus-midday share difference---our cleanest measure of retiming. We then test robustness through a leave-one-city-out exercise, dropping each city in turn and re-estimating the interaction on the retiming outcome. For an analysis by place category, we estimate the same specification separately for each amenity class, and decompose the aggregate retiming differential into within-type and between-type (compositional) components with a shift-share approach \cite{}; this separates differences in behaviour within the same place types from differences in the mix of place categories across cities. In effect, it asks how much of the aggregate difference would remain if cities had the same mix of amenities, and how much instead reflects different responses within the same categories.

To characterise the \emph{shape} of daily activity beyond simple retiming, we decompose each city's busyness profile across the day into Fourier harmonics. We fit the normalised hourly shares to a harmonic model with two terms, $\hat{y}(h) = a_0 + a_1\cos(\omega h) + b_1\sin(\omega h) + a_2\cos(2\omega h) + b_2\sin(2\omega h)$, where $\omega = 2\pi/24$. The first harmonic (period 24 hours) captures the dominant single-peaked daily rhythm; the second harmonic (period 12 hours) captures any tendency toward a bimodal pattern---an afternoon lull flanked by earlier and later surges. We define the \emph{Bactrian index} as the ratio of the second harmonic amplitude to the first, $B = \sqrt{a_2^2+b_2^2}\,/\,\sqrt{a_1^2+b_1^2}$. A value near zero indicates a single smooth hump; higher values indicate a pronounced two-humped profile resembling the silhouette of a Bactrian camel. We compute this index separately for each city's hot-day and mild-day profiles, so that the change in bimodality under heat can be compared across climate groups.

% \end{linenumbers}

% Bibliography (Nature-style, with DOI support) — style is set in wlscirep.cls
\bibliography{references}

\section*{Acknowledgments}
The work is supported by funding from the Alexander von Humboldt Foundation and the Federal Ministry of Education and Research (Bundesministerium für Bildung und Forschung) of Germany. Furthermore, the authors thank especially Soon-Gyo Jung (Qatar Computing Research Institute) for his work on collecting the Google Popular Times data used in this study.

\section*{Author contributions statement}
\textbf{A.R.} Conceptualization, methodology, analysis, writing; \textbf{I.W.} and \textbf{T.K.} writing, reviewing and editing.

\subsection*{Data and code availability}
Access to the busyness data used in the analysis is granted for academic and non-commercial purposes upon request to the authors.

\section*{Competing interests}
The authors declare no competing interests.

% =========================
% Embedded Supplementary Information
% =========================
\beginsupplementaryinformation

\begin{center}
{\LARGE Supplementary Information for\\[0.5\baselineskip]
\textbf{No one likes it hot, but hotter cities adjust by staying active later}\par}
\vspace{0.9\baselineskip}
{\large Andrew Renninger*, Till Koebe and Ingmar Weber \par}
\vspace{0.35\baselineskip}
{{\normalsize $^*$Corresponding author: Andrew Renninger (E-mail: renningera@ceu.edu)}\par}
\end{center}

% \supplementarytableofcontents
\clearpage

\section{Describing the data}

\begin{figure*}[ht!]
\centering
\includegraphics[width=1\textwidth]{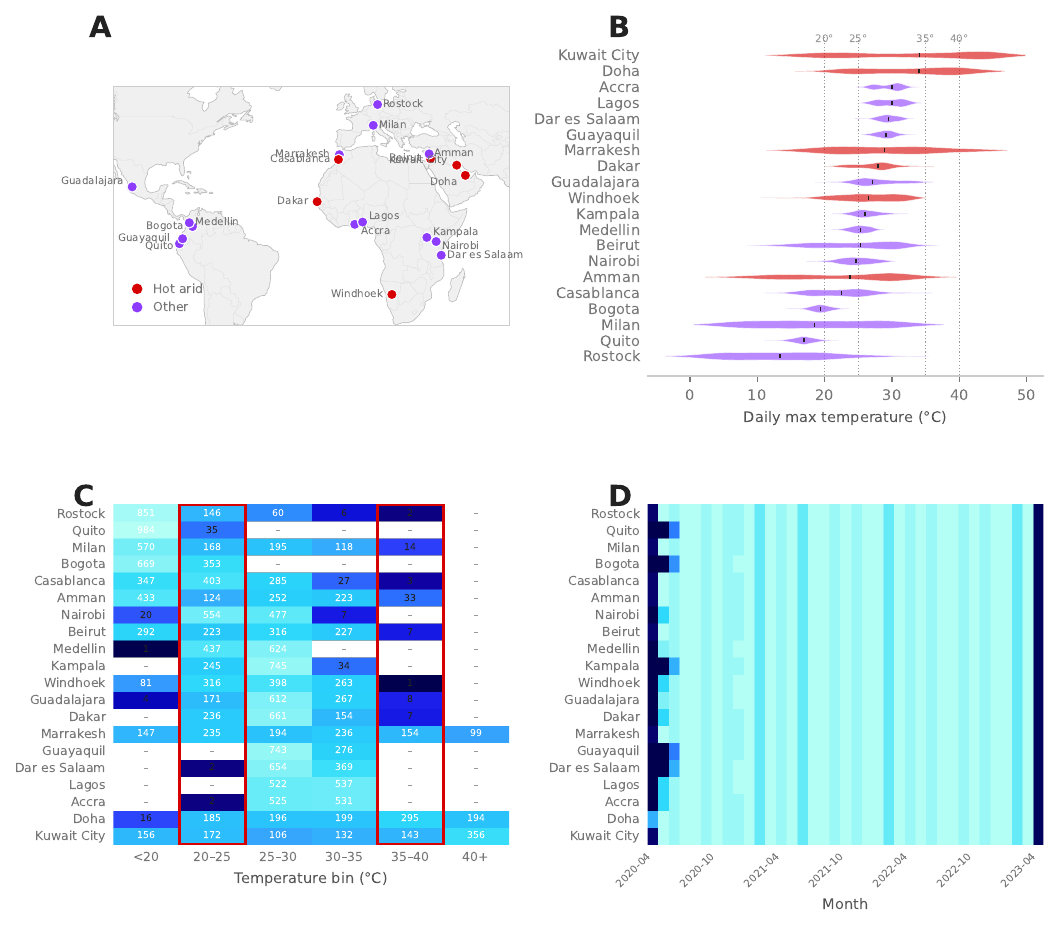}
\caption{\textbf{Sample geography, climate coverage, and observation window.}
\textbf{A}~Locations of the 20 cities in the sample, coloured by climate
classification: hot-arid (red) and other (purple); the sample spans four
continents, with hot-arid cities concentrated in the Middle East, North Africa,
and southern Africa. \textbf{B}~Distribution of daily maximum temperature across
all observed days for each city, sorted by median; dotted reference lines mark
the boundaries of the temperature bins used in estimation (20, 25, 35, and
40\textdegree C), showing that the 35--40\textdegree C contrast is well
supported in the hot-arid group but rare elsewhere. \textbf{C}~Number of
observed days per city and temperature bin; red outlines highlight the two bins
that form the core causal contrast (20--25\textdegree C and 35--40\textdegree C);
Doha and Kuwait City contribute the deepest support above 35\textdegree C, while
most non-hot-arid cities have no days in this range. \textbf{D}~Monthly
observation density by city, showing near-complete coverage from mid-2020 through
early 2023, with partial coverage in the first months of 2020 as cities entered
the scraping window and a truncation in April 2023.}
\label{fig:si_sample}
\end{figure*}

\clearpage

\begin{figure*}[ht!]
\centering
\includegraphics[width=1\textwidth]{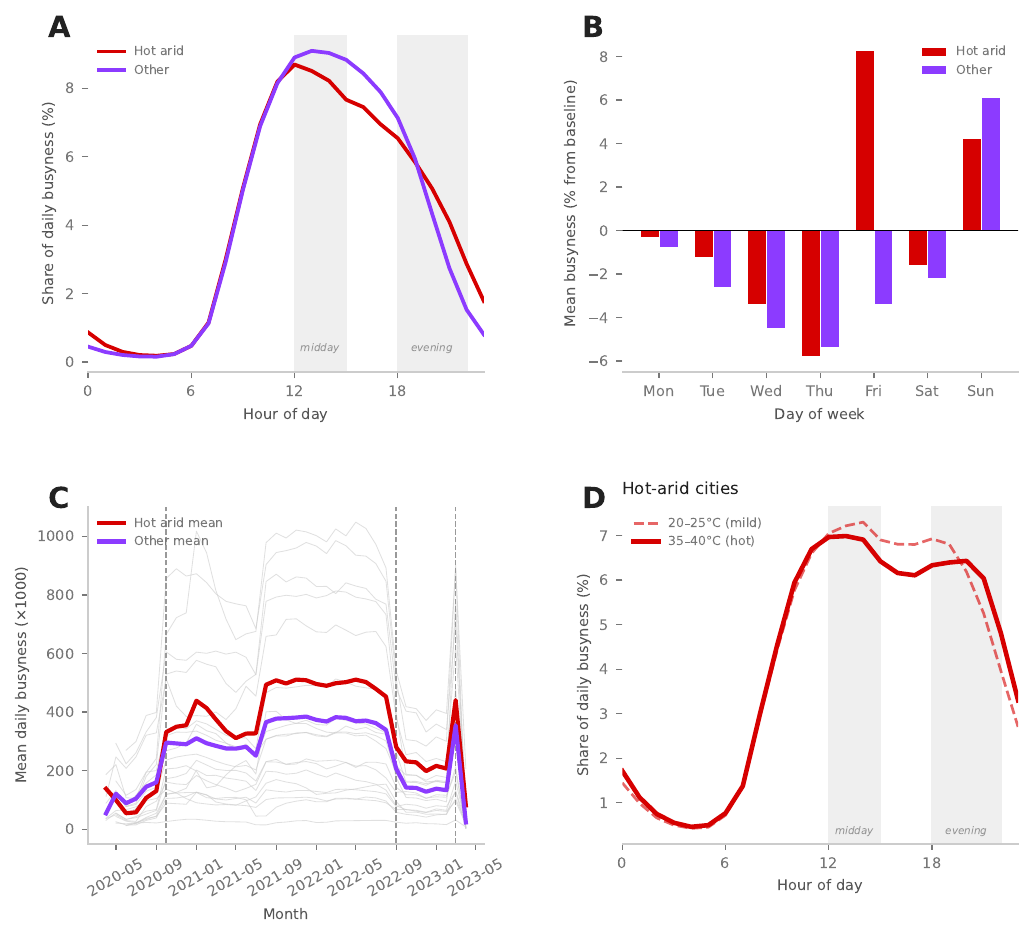}
\caption{\textbf{Properties of the busyness signal.}
\textbf{A}~Mean hourly share of daily activity across all city-days, split by
climate group, with shaded $\pm$1 standard error bands; both groups follow a
characteristic single-humped curve peaking in the early afternoon, though
hot-arid cities show a slightly wider plateau extending into the evening.
\textbf{B}~Mean daily busyness by day of week, expressed as percentage deviation
from each city's baseline; hot-arid cities show a pronounced Friday spike
reflecting the Islamic weekend in Gulf and Levantine cities, while other cities
peak on Sundays---both patterns are absorbed by the day-of-week fixed effects in
our specifications. \textbf{C}~Monthly mean raw busyness for individual cities
(grey) and group averages (bold), revealing at least three discrete rebasing
events (dashed lines) where Google recalibrated the Popular Times index; because
these level shifts are synchronous across cities, they are absorbed by the
year--month fixed effects, and the share-based retiming outcomes are mechanically
immune. \textbf{D}~The retiming effect visualised directly: hourly profiles for
hot-arid cities on days reaching 35--40\textdegree C (solid) versus
20--25\textdegree C (dashed), showing the characteristic midday flattening and
evening elevation that our TWFE estimates capture.}
\label{fig:si_signal}
\end{figure*}

\clearpage

\begin{table}[h!]
\centering
\caption{\textbf{City-year summary statistics.} Number of observed days, mean daily maximum temperature (\textdegree C), temperature range, days at or above 35\textdegree C, and mean daily precipitation (mm). Cities above the mid-rule are hot-arid. 2023 covers January--April only.}
\label{T1}
\scriptsize
\setlength{\tabcolsep}{6pt}
\begin{tabular}{@{}l rrrr rrrr rrrr rrrr @{}}
\toprule
 & \multicolumn{4}{c}{Days observed} & \multicolumn{4}{c}{Mean $T_{\max}$ (\textdegree C)} & \multicolumn{4}{c}{Days $\geq$35\textdegree C} & \multicolumn{4}{c}{Precip.\ (mm/day)}\\
\cmidrule(lr){2-5} \cmidrule(lr){6-9} \cmidrule(lr){10-13} \cmidrule(lr){14-17}
City & '20 & '21 & '22 & '23 & '20 & '21 & '22 & '23 & '20 & '21 & '22 & '23 & '20 & '21 & '22 & '23\\
\midrule
Amman         & 246 & 363 & 365 &  91 & 26.1 & 22.8 & 22.1 & 13.2 &  17 &   5 &  11 &  0 & 0.6 & 0.7 & 0.8 & 1.8\\
Dakar         & 239 & 363 & 365 &  91 & 28.4 & 27.0 & 27.6 & 25.6 &   3 &   0 &   4 &  0 & 1.5 & 0.6 & 1.3 & 0.2\\
Doha          & 266 & 363 & 365 &  91 & 35.4 & 33.3 & 32.8 & 24.2 & 161 & 166 & 162 &  0 & 0.2 & 0.0 & 0.1 & 1.5\\
Kuwait City   & 246 & 363 & 365 &  91 & 36.2 & 33.2 & 32.5 & 20.3 & 155 & 177 & 167 &  0 & 0.4 & 0.2 & 0.4 & 1.7\\
Marrakesh     & 246 & 363 & 365 &  91 & 31.4 & 28.2 & 29.3 & 22.6 &  89 &  73 &  89 &  2 & 0.5 & 0.6 & 0.7 & 0.9\\
Windhoek      & 240 & 363 & 365 &  91 & 26.0 & 26.0 & 26.2 & 28.9 &   0 &   1 &   0 &  0 & 0.3 & 1.3 & 1.0 & 2.0\\
\midrule
Accra         & 239 & 363 & 365 &  91 & 28.6 & 29.7 & 29.4 & 31.2 &   0 &   0 &   0 &  0 & 2.8 & 2.3 & 2.8 & 0.9\\
Beirut        & 246 & 363 & 365 &  91 & 27.1 & 24.6 & 23.8 & 18.0 &   5 &   2 &   0 &  0 & 0.9 & 2.3 & 1.8 & 3.8\\
Bogota        & 203 & 363 & 365 &  91 & 19.8 & 19.6 & 19.1 & 19.2 &   0 &   0 &   0 &  0 & 1.9 & 2.8 & 3.4 & 3.9\\
Casablanca    & 246 & 363 & 365 &  91 & 24.1 & 21.7 & 22.7 & 18.2 &   2 &   1 &   0 &  0 & 0.6 & 1.1 & 0.9 & 1.3\\
Dar es Salaam & 206 & 363 & 365 &  91 & 28.9 & 29.4 & 29.5 & 30.8 &   0 &   0 &   0 &  0 & 1.7 & 2.4 & 2.0 & 2.4\\
Guadalajara   & 243 & 363 & 365 &  91 & 28.0 & 27.4 & 28.1 & 28.1 &   2 &   1 &   5 &  0 & 1.9 & 2.0 & 1.7 & 0.0\\
Guayaquil     & 200 & 363 & 365 &  91 & 29.2 & 28.9 & 29.0 & 30.0 &   0 &   0 &   0 &  0 & 0.6 & 3.1 & 3.1 &12.6\\
Kampala       & 205 & 363 & 365 &  91 & 25.4 & 26.1 & 26.4 & 27.6 &   0 &   0 &   0 &  0 & 3.8 & 3.4 & 3.2 & 2.9\\
Lagos         & 240 & 363 & 365 &  91 & 28.9 & 30.1 & 29.7 & 31.9 &   0 &   0 &   0 &  0 & 4.6 & 3.3 & 3.3 & 1.2\\
Medellin      & 243 & 363 & 365 &  91 & 25.5 & 25.2 & 25.0 & 25.4 &   0 &   0 &   0 &  0 & 6.7 & 6.2 & 7.4 & 5.4\\
Milan         & 246 & 363 & 365 &  91 & 21.1 & 18.0 & 20.0 & 12.1 &   2 &   0 &  12 &  0 & 4.2 & 2.8 & 2.2 & 1.2\\
Nairobi       & 239 & 363 & 365 &  91 & 23.7 & 24.6 & 25.0 & 27.3 &   0 &   0 &   0 &  0 & 2.2 & 2.2 & 1.2 & 0.9\\
Quito         & 200 & 363 & 365 &  91 & 17.8 & 16.9 & 16.7 & 16.2 &   0 &   0 &   0 &  0 & 6.5 &11.5 &10.8 &16.7\\
Rostock       & 246 & 363 & 365 &  91 & 15.9 & 12.5 & 13.6 &  6.9 &   0 &   0 &   2 &  0 & 2.2 & 2.2 & 1.8 & 2.4\\
\bottomrule
\end{tabular}
\end{table}

\clearpage

\section{Robustness checks}

\begin{figure*}[ht!]
\centering
\includegraphics[width=1\textwidth]{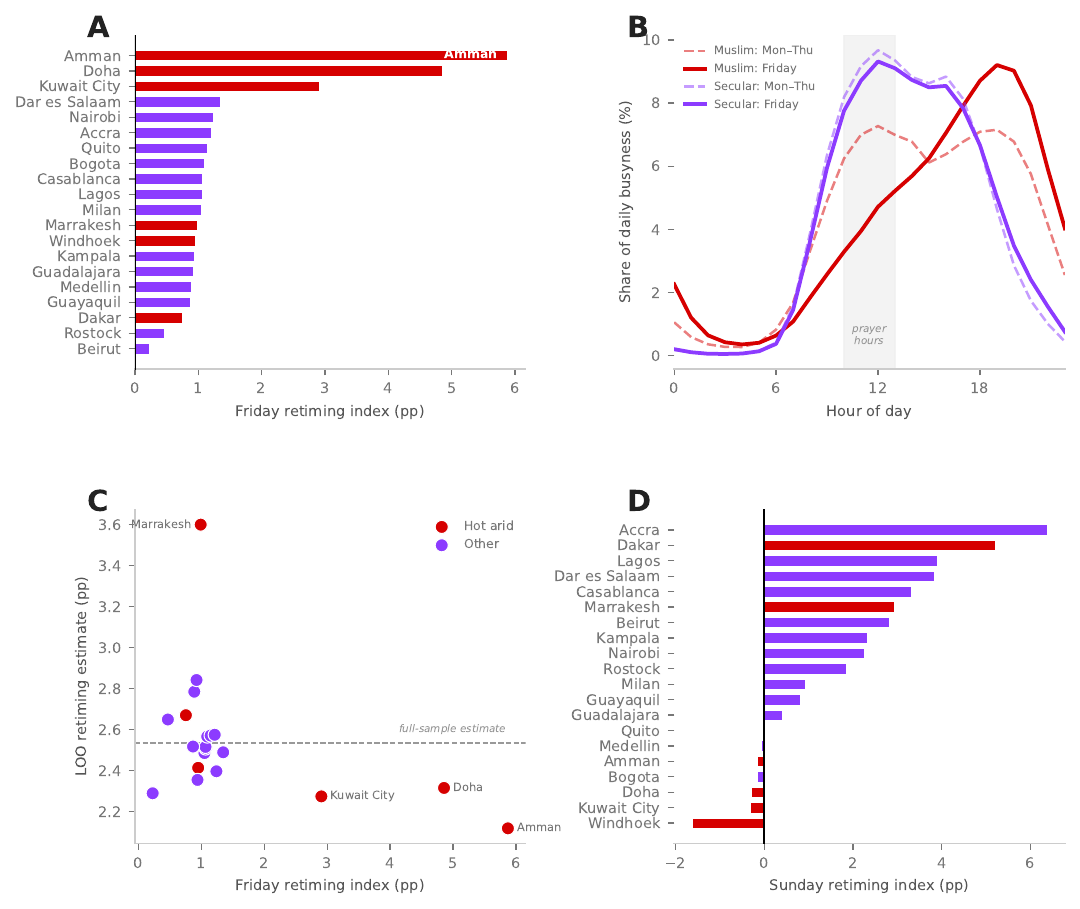}
\caption{\textbf{Day-of-week cultural rhythms do not drive the retiming result.}
\textbf{A}~Friday retiming index---the shift in activity from midday to evening
on Fridays relative to Monday--Thursday---ranked by city; Amman, Doha, and Kuwait
City show large Friday prayer effects, while the remaining hot-arid cities
(Marrakesh, Dakar, Windhoek) are indistinguishable from secular cities.
\textbf{B}~Hourly profiles on Fridays (solid) versus weekdays (dashed) for
Muslim-majority cities (red) and secular controls (purple); the midday crater
and evening surge in Muslim cities mirror the shape of the heat-retiming effect,
motivating the concern that religious calendars could confound our estimates.
\textbf{C}~Leave-one-city-out retiming estimates plotted against each city's
Friday index; cities with the strongest prayer effects (Amman, Doha) produce the
lowest leave-one-out estimates, but dropping Amman---the most extreme
case---still leaves the retiming differential at +2.1~pp, and there is no
positive association between Friday strength and the estimated heat effect.
\textbf{D}~The same index computed for Sundays, showing that rest-day retiming
is universal across all cities regardless of religious tradition; the
day-of-week fixed effects in our specification absorb these cultural rhythms
symmetrically.}
\label{fig:friday_robustness}
\end{figure*}

\clearpage

\begin{figure*}[ht!]
\centering
\includegraphics[width=1\textwidth]{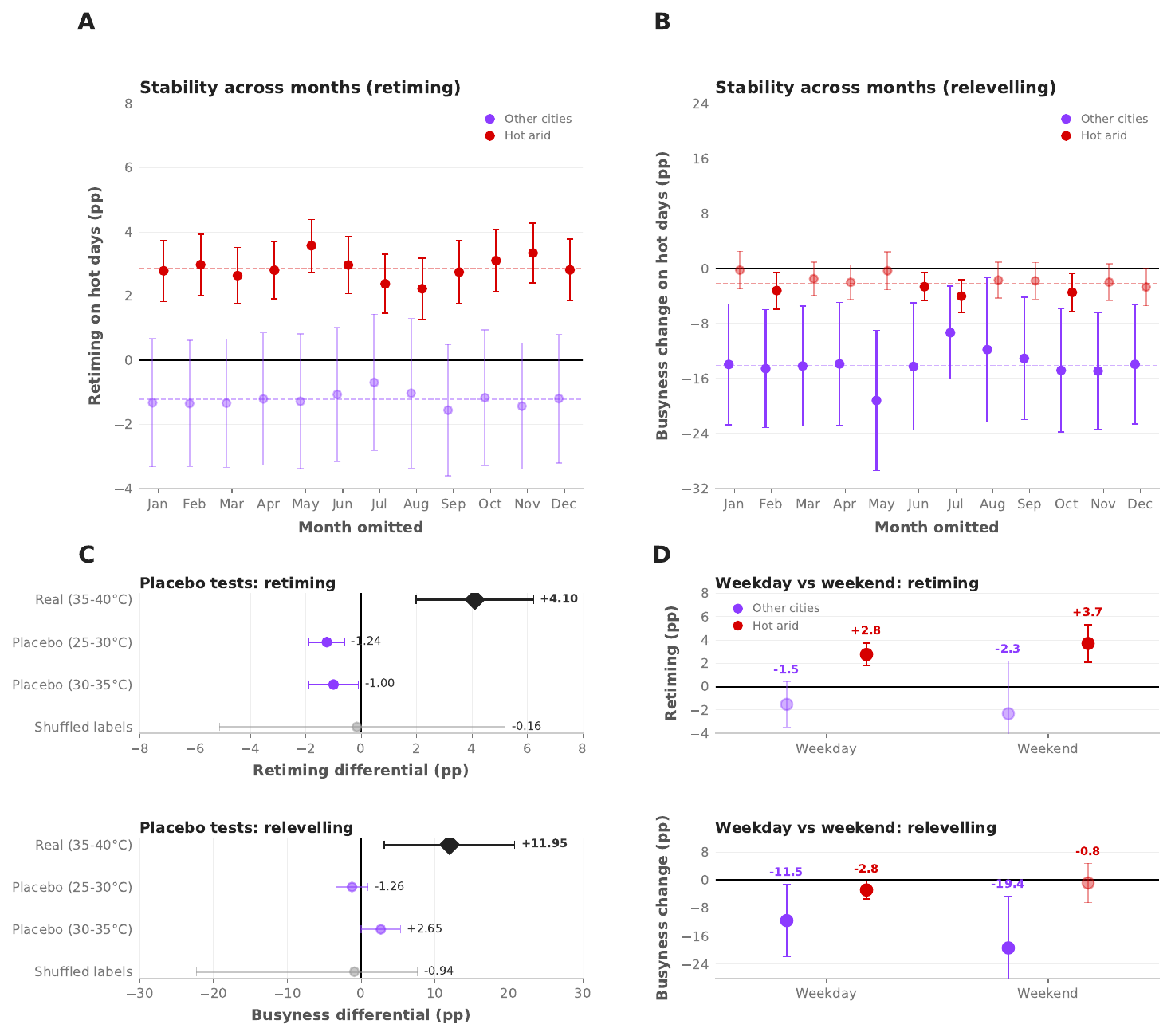}
\caption{\textbf{Retiming and relevelling in hot, arid cities are robust.} \textbf{A} Leave-one-month-out estimates of the retiming response, dropping each calendar month in turn; hot, arid cities stay consistently above other cities across every month dropped, with July and August producing the lowest but still-positive estimates, ruling out the concern that summer population outflows are driving the gap. \textbf{B} The same leave-one-month-out exercise for relevelling (total busyness change), showing that other cities lose around 15 percentage points of activity on hot days while hot, arid cities lose almost none, and this gap persists regardless of which month is omitted. \textbf{C} Placebo forest plot showing the real effect (35--40\textdegree C vs.\ 20--25\textdegree C) against three placebo specifications: cooler temperature contrasts (25--30\textdegree C and 30--35\textdegree C vs.\ 20--25\textdegree C) yield null or slightly negative differentials, indicating that the retiming response is specific to extreme heat rather than a generalised temperature sensitivity; randomisation inference from 500 shuffled hot-arid labels produces a null distribution centred near zero with the real estimate sitting in the upper tail. \textbf{D} Splitting the sample by day type, we show that retiming and relevelling gaps hold on both weekdays and weekends, with weekend relevelling gaps actually widening---hot, arid cities lose almost no activity on hot weekends while other cities lose nearly 20 percentage points---indicating the response is not merely a product of after-work leisure time.}
\label{fig:si_robustness}
\end{figure*}

\clearpage

\begin{figure*}[ht!]
\centering
\includegraphics[width=1\textwidth]{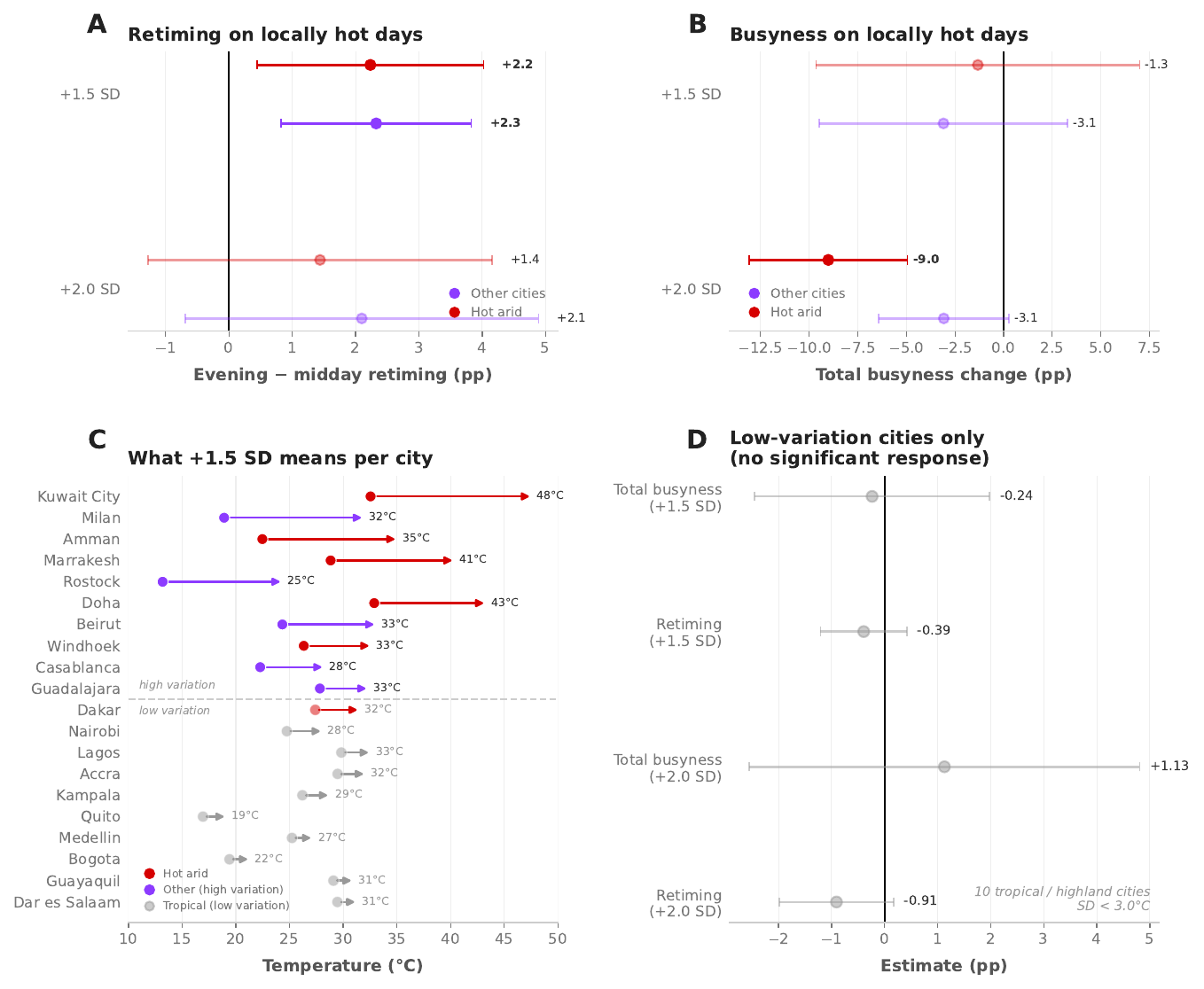}
\caption{\textbf{Retiming is robust to locally defined heat thresholds, but only in cities with temperature variation.} \textbf{A} Retiming estimates using within-city SD-based thresholds (+1.5 and +2.0 SD above the city mean), restricted to cities with SD \textgreater{} 3.0\textdegree C; both hot, arid and other cities show significant retiming on locally hot days, with hot, arid cities retiming somewhat more at the +2.0 SD threshold. \textbf{B} Total busyness effects at the same thresholds; hot, arid cities see large activity losses on locally hot days, while other cities show smaller and noisier declines. \textbf{C} What +1.5 SD means in absolute terms for each city, illustrating why we restrict the sample: tropical and highland cities have temperature SDs below 3\textdegree C, so +1.5 SD corresponds to excursions of only 2--3\textdegree C, while cities with real seasonality see excursions of 5--15\textdegree C. \textbf{D} Running the same specification on the low-variation cities confirms the absence of any retiming or destruction signal, consistent with these cities rarely experiencing meaningfully anomalous temperatures.}
\label{fig:si_sd_variation}
\end{figure*}

\clearpage

\begin{figure*}[ht!]
\centering
\includegraphics[width=1\textwidth]{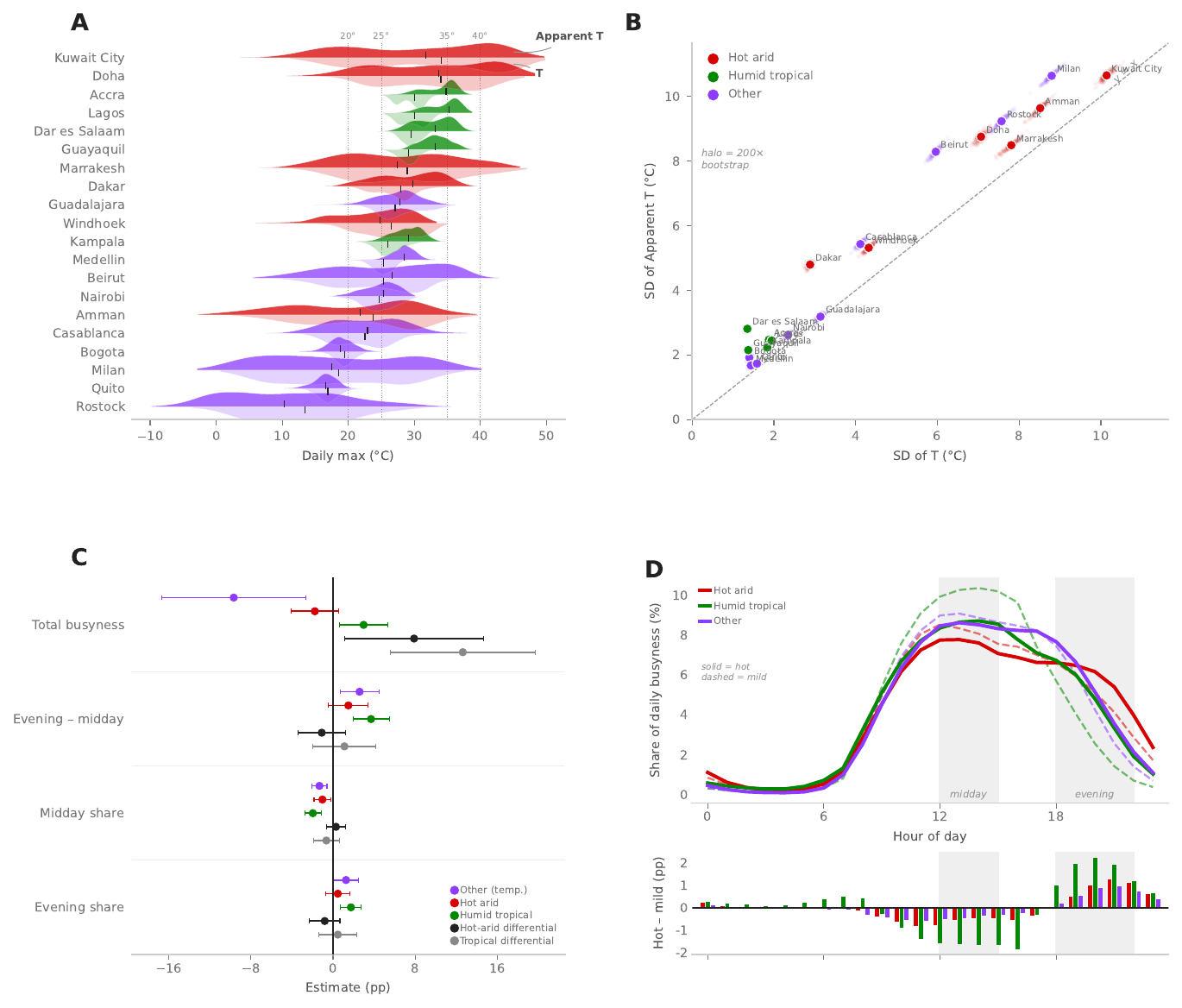}
\caption{\textbf{Apparent temperature widens the variation and reveals tropical retiming.} \textbf{A} Distributions of daily maximum temperature for each city, with dry-bulb below each city's row and apparent temperature above; humidity shifts humid tropical and coastal cities several degrees rightward, pushing many into the 30--40\textdegree C range where our causal contrast lives. \textbf{B} Standard deviation of apparent temperature against the equivalent for absolute values, with 200-iteration paired bootstrap halos conveying estimation uncertainty; nearly every city sits at or above the identity line, meaning apparent temperature generally adds day-to-day variation, with the largest absolute gains in cities that already had meaningful seasonality. \textbf{C} Two-way fixed-effects estimates on 35--40\textdegree C versus 20--25\textdegree C apparent days, split into hot-arid, humid tropical, and temperate groups, with the two hot-arid and humid-tropical differentials against the temperate reference; point estimates and signs align with the main text---temperate cities lose activity, hot-arid cities hold steady, humid tropical cities hold or gain---but statistical significance weakens because splitting the non-hot-arid pool into two strata leaves each coefficient identified from fewer city clusters, inflating cluster-robust standard errors. \textbf{D} Hourly activity profiles on 35--40\textdegree C apparent (solid) versus 20--25\textdegree C apparent (dashed) days with hot-minus-mild differences below; humid tropical cities show the sharpest retiming---the deepest midday depression and the largest evening gain---a pattern the absolute contrast could not reveal because these cities rarely reach 35\textdegree C on the thermometer.}
\label{fig:si_apparent}
\end{figure*}

\clearpage

\section{Hourly profiles}

\begin{figure*}[ht!]
\centering
\includegraphics[width=1\textwidth]{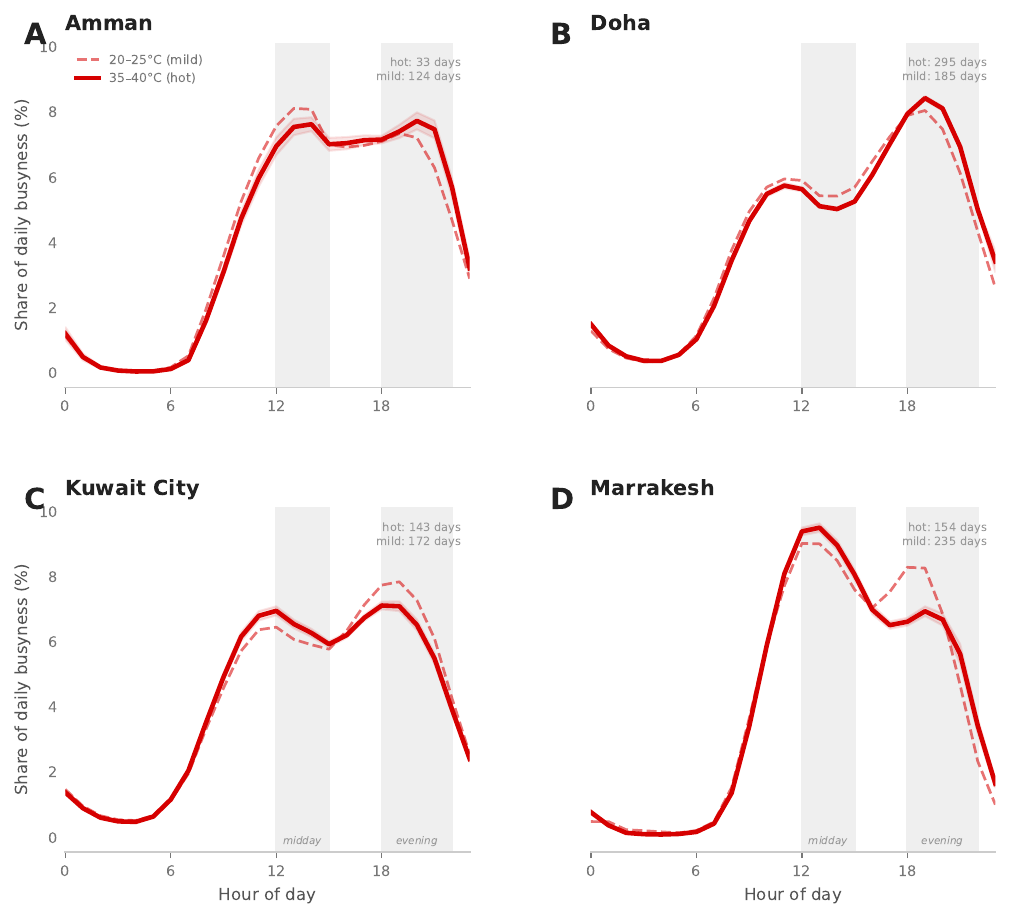}
\caption{\textbf{Hourly activity profiles on hot and mild days for four
hot, arid cities.} Each panel shows the mean
share of daily busyness by hour on days reaching 35--40\textdegree C (solid,
with $\pm$1 SE shading) versus 20--25\textdegree C (dashed), with the number of
contributing days noted. \textbf{A}~Amman shows modest retiming with a slightly
flattened midday. \textbf{B} ~Doha's hot-day profile collapses into a single
evening peak, explaining its negative Bactrian shift: the distribution does not
become more bimodal, it concentrates into the evening. \textbf{C}~Kuwait City
shows a similar pattern to Doha, with the midday plateau suppressed and evening
activity sustained. \textbf{D} ~Marrakesh exhibits the most pronounced Bactrian
shift in the sample: on hot days, the midday trough deepens sharply, producing a
two-humped profile with distinct morning and evening peaks.}
\label{fig:city_profiles}
\end{figure*}

\clearpage

\section{Composition effects}

\begin{figure*}[ht!]
\centering
\includegraphics[width=1\textwidth]{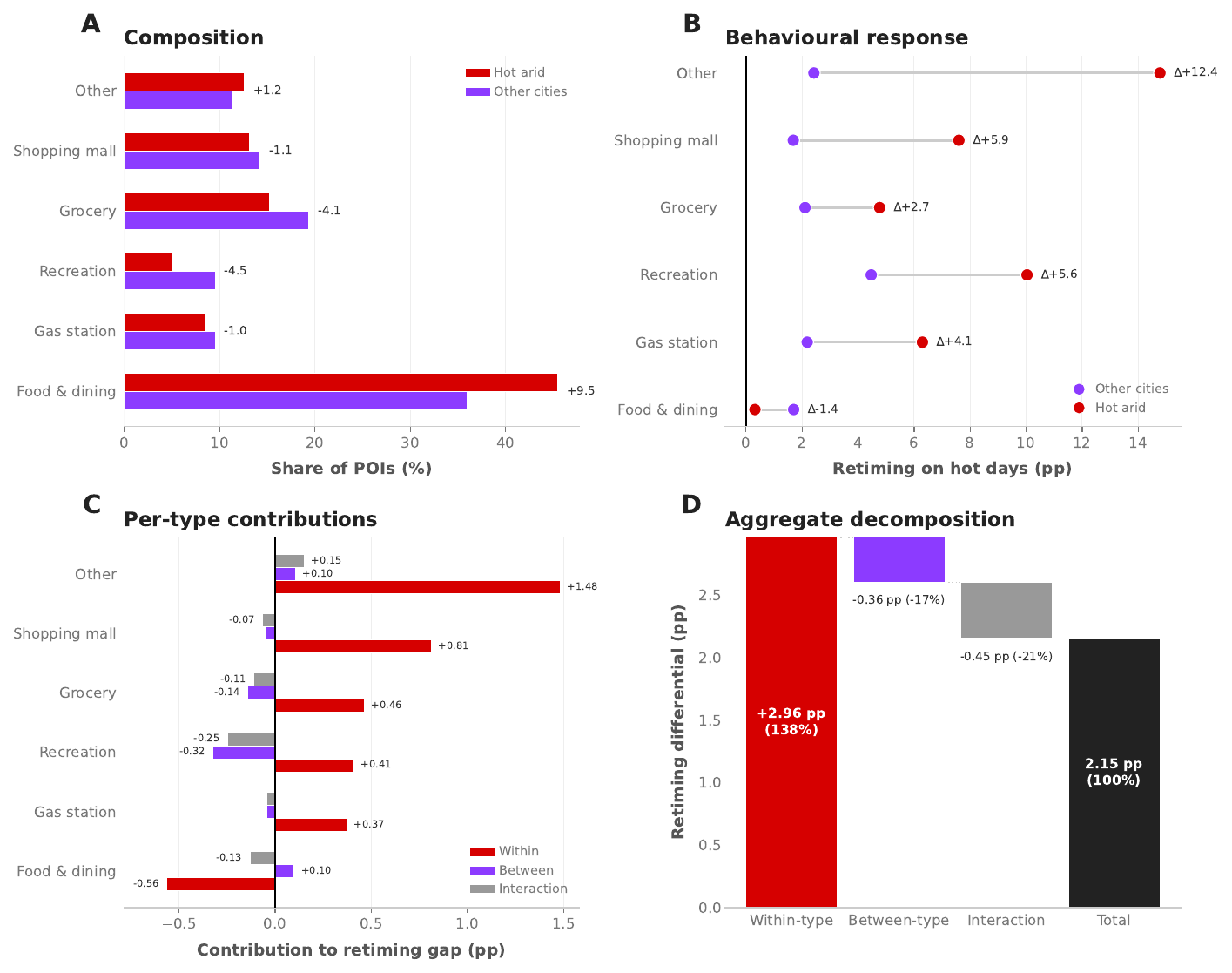}
\caption{\textbf{The retiming gap between climate zones is driven by behavioural differences within categories, not composition.} \textbf{A} Baseline composition of place types in hot, arid cities versus others; hot, arid cities are disproportionately food-heavy and have fewer recreation venues. \textbf{B} Retiming responses to extreme heat by place type, showing that nearly all categories retime more in hot, arid cities, with especially large gaps for other commercial venues and recreation. \textbf{C} Per-type contributions to the aggregate retiming gap, decomposed into within-type (same mix, different behaviour), between-type (different mix, same behaviour), and interaction terms; the within-type channel dominates, while between-type and interaction contributions largely cancel across categories. \textbf{D} Aggregating across place types, within-type behavioural differences account for 138\% of the retiming gap, partially offset by small negative between-type and interaction terms that reflect offsetting composition differences.}
\label{fig:si_shiftshare}
\end{figure*}

\clearpage

\end{document}